# Planes, Chains, and Orbits: Quantum Oscillations and High Magnetic Field Heat Capacity in Underdoped YBCO

Scott C. Riggs[1*]; O. Vafek[1], J. B. Kemper[1], J.B. Betts[2], A. Migliori[2], W. N. Hardy[3,4], Ruixing Liang[3,4], D. A. Bonn[3,4], G.S. Boebinger[1]

[1]National High Magnetic Field Laboratory, Florida State University, Tallahassee-FL 32310, USA
[2]Los Alamos National Laboratory, Los Alamos, NM 87545, USA
[3]Department of Physics and Astronomy, University of British Columbia, Vancouver, BC, V6T 1Z1, Canada
[4]Canadian Institute for Advanced Research, Toronto, Canada
[*]To whom correspondence should be addressed; E-mail: scr9035@stanford.edu

**The underlying physics of the magnetic-field-induced resistive state in high temperature cuprate superconductors remains a mystery. One interpretation is that the application of magnetic field destroys the d-wave superconducting gap to uncover a Fermi surface that behaves like a conventional (i.e.Fermi Liquid) metal (*1*). Another view is that an applied magnetic field destroys long range superconducting phase coherence, but the superconducting gap amplitude survives (*2, 3*). By measuring the specific heat of ultra-clean $YBa_2Cu_3O_{6.56}$ (YBCO 6.56), we obtain a measure of the quasi-particle density of states from the superconducting state well into the magnetic-field-induced resistive state. We have found that at very high magnetic fields the specific heat exhibits both the conventional temperature dependence and quantum oscillations expected for a Fermi Liquid. On the other hand, the magnetic field dependence of the quasi-particle density of states follows a $\sqrt{H}$ behavior that persists right through the zero-resistance transition, evidencing the fully developed d-wave superconducting gap over the entire magnetic field range measured. The coexistence of these two phenomena pose a rigorous thermodynamic constraint on theories of high-magnetic-field resistive state in the cuprates.**

Specific heat of a material is a measure of heat necessary to raise the temperature of a given amount of material, typically a gram or a mol, by 1 Kelvin. Near absolute zero, this bulk thermodynamic quantity is a sensitive probe of the low energy excitations of a complex quantum system. Such low energy excitations contain useful information about the nature of the ground state. For the canonical example of an ordinary (so-called 'Fermi liquid') metal, the electronic (charge-carrying quasiparticles) component of specific heat vanishes linearly with temperature and the lattice vibration (phonon) component vanishes as $T^3$ . The two different power-laws ultimately arise from the different quantum statistics of electrons (fermions) and phonons (bosons). With the application of an intense magnetic field $H$, the electronic orbits are quantized in a Fermi liquid metal, resulting in oscillations that are periodic in $1=H$ in many physical quantities, including magnetoresistance, magnetization, and specific heat. In superconductors, specific heat probes the low energy excitations governed by the magnitude and symmetry of the superconducting energy gap.

We report the specific heat of an underdoped, high-quality, high temperature superconductor YBCO 6.56 as a function of temperature and magnetic field up to $45T$ . Samples in this doping range are of particular interest, as the level of disorder is small enough for the observation of quantum oscillations in the resistive state (*1*) (*4*) (*5*) (*6*). One of the motivations of the specific heat measurements reported here is to probe the evolution of the system's low energy excitations with increasing magnetic field from the nodal d-wave superconducting state well into the high-field resistive state. The nature of this, field-induced, resistive state in underdoped cuprates at low temperatures is a subject of intense attention (*2, 3,7–11*) that is undergoing a resurgence (*1*) (*4*) (*5*) (*6*) (*12*) (*13*) (*14*) because understanding the resistive state is essential to understanding high temperature cuprate superconductivity, arguably the greatest challenge in condensed matter physics today.

At low $T$ and magnetic field $H$ of order a few Tesla -too small to suppress superconductivity- it is established that the system is a d-wave superconductor in the vortex state (*15*) (*16*). Previous measurements on optimally doped YBCO (*17*) (*18*) (*19*) (*20*) performed at magnetic fields up to ~$16T$ observed a low temperature $T$ -linear coefficient to specific heat, $\gamma(H)$, that grows as $A_c\sqrt{H}$. Such $H$-dependence is understood to be a signature of a d-wave energy gap at low magnetic fields (*21*), where the energies associated with Landau-level quantization, $\omega = \frac{\hbar eH}{m^* c}$, and the d-wave gap $\Delta_0$, satisfy the relation $\hbar\omega_c \ll \Delta_0$. Here $m_$ is the effective mass and $c$ is the speed of light. In this low-magnetic-field regime, many Landau levels are thoroughly mixed by the pairing term (*22*) (*23*). At the semiclassical level, the $\sqrt{H}$ increase of the electronic contribution to the specific heat arises from the Doppler shift of d-wave quasiparticles in the vicinity of vortices (*24*) (*25*). It is the magnetic-field-induced superconducting vortices which give rise to the additional low lying quasiparticles that contribute to the specific heat. Heat capacity measurements in magnetic fields up to ~16*T* probe the nature of the superconducting state, but are unable to access the resistive state at low temperatures (*17*) (*18*) (*19*). Here we report specific heat measurements in YBCO over the magnetic field range from 0 to 45*T* , spanning the irreversibility field ($H_{irr}$ ~23±4*T* at 1.5*K*) above which vortex motion causes the superconductor to become resistive (*26*) (*27*). Only such high fields enable the tracking of low energy electronic excitations *continuously* through the resistive transition. Our results in Figure 1 show two distinct contributions to the electronic specific heat: a high-field oscillatory component atop a $\sqrt{H}$ background that extends *unperturbed* over the entire magnetic field range studied (*28*). Figure 1(b) shows the $\sqrt{H}$ background with the oscillatory component subtracted, as described below. Figure 1(c) contains the data of figure 1(a) with the $\sqrt{H}$ component subtracted.

The persistence of a simple $\sqrt{H}$ behavior across $H_{irr}$ means that the thermodynamic d-wave energy gap is unaffected by the resistive transition, where $A_c$, the prefactor of the $\sqrt{H}$ contribution, is a measure of the Dirac cone anisotropy at the nodes in the d-wave energy gap. Previously-reported thermal conductivity experiments (*29*) estimate a factor of two increase of the Dirac cone anisotropy from ≈7.9 for underdoped YBCO 6.54 to ≈15:5 for optimally-doped YBCO 6.99. We find the same ratio in specific heat: $A_c$~0.47 $mJ\ mol^{-1}\ K^{-2}\ T^{-0.5}$ that we measure in underdoped YBCO 6.56 is roughly half the value (0.88 - 1.3 $mJ\ mol^{-1}\ K^{-2}\ T^{-0.5}$ ) obtained from the low-field measurements in optimally doped samples (*17*) (*19*). Thus both the functional form *and magnitude* of the $\sqrt{H}$ dependence indicate a fully developed d-wave superconducting gap at all magnetic fields to 45T *even in the resisitve state*. From the agreement with the semiclassical approximation (*24*) at $H = 45T$, it appears that we remain in the low-field limit of the still-robust superconducting phase, even though the system is fully-resistive.

The high-field oscillatory contribution, visible at fields above $H_{irr}$ in Figure 1, is isolated and plotted in the Figure 2 for closer study. Similar quantum oscillations have been reported in YBCO from magneto-transport and magnetization measurements (*1*) (*4*) (*5*) (*6*). Specific heat measurements provide an important advantage in that they probe the nature of all low energy excitations *continuously* across $H_{irr}$ , below which both transport and magnetization signals are overwhelmingly dominated by the superconducting condensate. In a clean quasi two-dimensional Fermi Liquid the oscillatory component of the specific heat due to Landau level quantization of the orbits is given by the Lifshitz-Kosevich (LK) formula

$$\Delta C_v(T,H) = -AR_D T \sum_{p=1}^{\infty} J_0\left(4\pi p \frac{t_\omega}{\hbar\omega_c}\right) \cos\left(2\pi p\left(\frac{\mu}{\hbar\omega_c} - \frac{1}{2}\right)\right) f''(x) \qquad (1)$$

where $A$ is a constant, $R_D = \exp\left(\frac{-Cm^*T_D}{H}\right)$ is the Dingle term where $C$ is a constant, $T_D$ is the Dingle temperature, $\hbar\omega_c = \frac{\hbar eH}{m^*C}$, $x = \frac{2\pi^2 p k_B T}{\hbar\omega_c}$, $f''(x) = x\frac{1+cosh^2(z)}{sinh^3(z)} - \frac{2\cosh(z)}{sinh^2(z)}$, $J_0$ is a Bessel function of the first kind a $t_w$ is the c-axis hopping energy, resulting in a small warping of the two dimensional Fermi surface. The frequency of the oscillations is determined by the cross-sectional area of the Fermi surface or equivalently by the chemical potential $\mu$. Quantum oscillation transport measurements find a Dingle temperature of 6\.4$K$ for YBCO 6.59 (*30*). We note that, even if the Dingle temperature were as high as 12$K$ in our sample, the variation of the Dingle term (*31*) would be negligible in magnetic field (30$T$-45$T$) and temperature (1$K$-5.5$K$) ranges over which we measure specific heat quantum oscillations. The oscillations shown in Figure 2 were measured while sweeping the magnetic field. We fit our data between 30-45 T to the first harmonic (p=1 term) of the LK formula. Note that the function $f''(x)$ changes sign near $z$ = 1.6, which accounts for the $\pi$-phase shift that is clearly visible in our data between 1.7$K$ and 5.5$K$ (dashed line in Figure 2(B)). The $f''(x) = 0$ node determines an effective mass $m^* \approx 1.34 \pm 0{:}06\ me$. The frequency of the oscillations F=531 ± 3T, the warping term $t_w$ = 15.2$T$ and the amplitude $A = 1.39 \pm 0.4\ mJ\ mol^{-1}\ K^{-2}$ are fitted to our $T$ = 1$K$ data, for which the signal to noise ratio is the highest. The resulting formula is then compared with the data at higher temperature without further fitting, as shown in Figure 2.

The detailed agreement of specific heat data with LK formalism adds to the compelling evidence in the literature (*1*) (*26*) (*27*) that conventional Fermi Liquid quasiparticles exist at high magnetic fields in underdoped YBCO. However, the specific heat data demonstrate for the first time that these LK quantum oscillations coexist with the $\sqrt{H}$ signature of the d-wave superconducting gap in the resistive regime. We now explore the surprising coexistence of the signatures of a Fermi Liquid and a d-wave superconductor within each of the two leading scenarios for the resistive transition: (a) that the application of magnetic field destroys the d-wave superconducting gap to uncover a Fermi surface that behaves like a conventional (i.e. Fermi Liquid) metal (*1*), and (b) that the applied magnetic field destroys long range superconducting phase coherence and the system enters a vortex liquid state (*4*).

The magnitude of the electronic specific heat can distinguish between these two scenarios. Figure 3 shows the linear temperature dependence of $C_v = T$ versus $T^2$ at $H = 0$ and $H = 45T$, such that $C_v/T = \gamma + \beta T^2$. The value of $\beta$, i.e. the $T^3$ contribution to $C_v$ originates from phonons and is naturally H independent. The finite value of $\gamma$ indicates a finite electronic density of states at low energy, even in zero field where our value $\gamma(H = 0) \approx 1.85$ $mJ\,mol^{-1}\,K^{-2}$ is comparable to the value obtained in clean optimally-doped $Y\,Ba_2Cu_3O_7$ samples (*20*) (*32*). It has been proposed that this large value for $\gamma(0)$ originates from a disorder-generated finite density of quasiparticle states near the d-wave nodes; however, this value in YBCO is larger than reported for LSCO, a material that is considered to be significantly more disordered (*33*).

For a two dimensional Fermi liquid with parabolic bands, in which the density of states is constant, the total electronic contribution to the specific heat is a simple sum over each piece -or pocket- of the Fermi surface:

$$\gamma_{total} = \gamma' \sum_i n_i\, m_i^* \quad (2)$$

where n is the number of each type of pocket per unit cell, $m^*$ is the effective mass of that pocket and $\gamma' = 1.46$ $mJ\,mol^{-1}\,K^{-2}$. In the scenario in which the high-field resistive state is interpreted as a normal state Fermi liquid, a single pocket per unit cell ($n = 1;\ m^* = 1.35m_e$) would yield $\gamma_{total} = 1.9$ $mJ\,mol^{-1}\,K^{-2}$ and a single pocket *per* $CuO_2$ *plane* (that is $n = 2;\ m^* = 1{:}35m_e$) would yield $\gamma_{total} = 3.8$ $mJ\,mol^{-1}\,K^{-2}$. The symmetry of the Brillouin zone (BZ) admits only two possibilities for a single pocket, each one shown in the upper left of Figure 3. It has been proposed that the high-field resistive state in underdoped YBCO is an anti-ferromagnetically-driven Fermi surface reconstruction (lower right inset of Figure 3), where there is one pocket per $CuO_2$ plane centered at $(\pm\pi;\ 0)(0;\pm\pi)$ with $m^* = 1.35m_e$ and two pockets at $(\pi/2;\pi/2)$ with $m^* = 3.8m_e$ (*4*) which would yield $\gamma_{total} = 26$ $mJ\,mol^{-1}\,K^{-2}$. Taken together, the small observed value of $\gamma \sim 2$ $mJ\,mol^{-1}\,K^{-2}$ from figure (1) and the persistence of the $\sqrt{H}$ signature of a d-wave superconducting gap effectively rule out the first scenario: i.e. the scenario in which high magnetic-field-resistive state destroys the d-wave superconducting gap to reveal a conventional normal state Fermi liquid.

Thus, we turn attention to the second scenario in which the high-magnetic-field state is a vortex liquid. A vortex liquid features the coexistence of Fermi liquid quasiparticles and mobile superconducting vortices, and *prima facia* would appear to be the most logical interpretation of our data. In a vortex liquid however, both the amplitude of the quantum oscillations and the value of γ in the electronic specific heat would be expected to be suppressed (*34*). The suppression of the quantum oscillations, ascribed in part to quasiparticle scattering from the vortices, would be expected to vary with the number of vortices, i.e. the applied magnetic field. A remarkable finding from all measurements of quantum oscillations in underdoped YBCO is that the standard LK formula (see equation 1) can fit all data, and thus, there is no evidence of a magnetic field damping term arising from vortex scattering up to magnetic fields as high as 85*T* (*26*).

The suppression of γ on the other hand might be ascribed to the volume fraction of the normal state in the sample. From the value calculated by pocket counting, gamma would need to be suppressed by roughly an order of magnitude to agree with our data in Figure 1(C). Remarkably, however, this suppression would also appear to be magnetic field independent from 45*T* down through the resistive transition at *Hirr* 23*T and continuing to zero magnetic field* (see Figure 3(C)).

Coexistence of a Fermi surface with the robust d-wave pairing gap suggests, *prima facie*, the existence of at least two distinct closed Fermi surfaces, one of which is d-wave gapped while the other is not. We, thus, venture a third scenario that proposes real-space separation of two weakly-coupled physical systems: It ascribes the $\sqrt{H}$ dependence in Figure 1(B) to a Fermi surface that is d-wave gapped at all magnetic fields to 45*T*. It interprets the data in Figure 1(C) in terms of a second Fermi surface, in which the quantum oscillations at high magnetic field arise from the same carriers that give the large zero field value of $\gamma \sim 2mJmol^{-1}K^{-2}$, which suggests that the value of $2mJmol^{-1}K^{-2}$ is intrinsic. A value of $\gamma \sim 2mJmol^{-1}K^{-2}$ corresponds to a single pocket per unit cell per equation 2.

Every aspect of the specific heat data are consistent with a *single* Fermi surface pocket in the unit cell, which behaves as a normal metal. Any other scenario would seem to require a detailed balance of purported deviations from the $\sqrt{H}$ dependence in Fig 1(B) at high magnetic fields to be precisely offset by the onset of a magnetic-field-induced specific heat component that ultimately gives rise to the quantum oscillations. There is little evidence for this in the specific heat data and no obvious physical reason that such a precise balance should exist. Most importantly, our measurements place a clear and strong constraint on any thermodynamic theory of the magnetic-field-induced resistive state in underdoped YBCO.

37. The authors gratefully acknowledge discussions with N. Harrison, P. Hirschfeld, S. Kivelson, P.A. Lee, R. McDonald, S. Sachdev, Z. Tesanovic, S. Todadri, C.M. Varma. The authors would also like to acknowledge F.F. Balakirev for programming of the data acquisition software. SCR acknowledges financial support from ICAM. WH, RL, DB are supported by the Natural Science and Engineering Research Council of Canada and the Canadian Institute for Advanced Research, The NHMFL is supported by the State of Florida and the National Science Foundation's Division of Materials Research through DMR-0654118 .

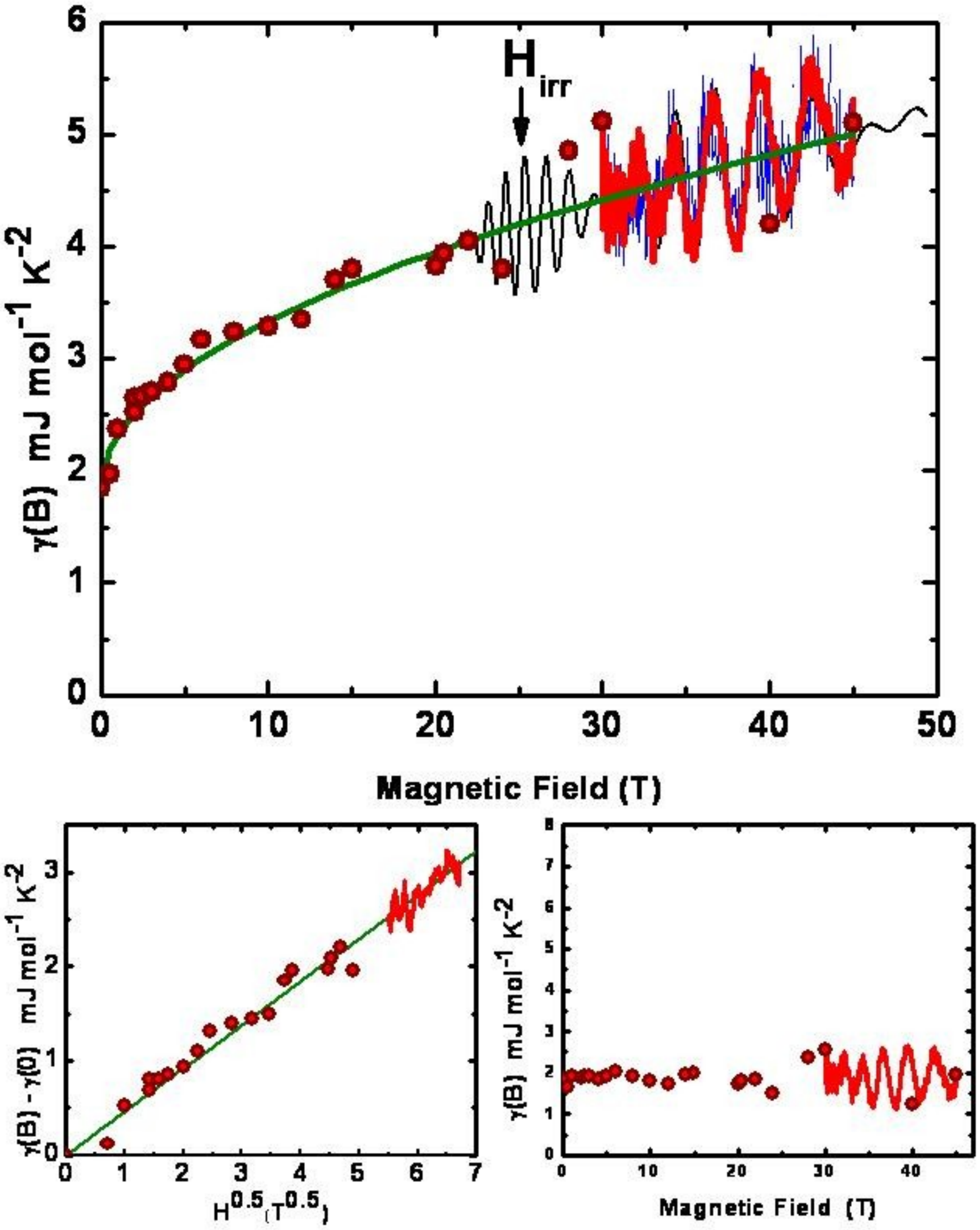

Figure 1: (A) Electronic component of the specific heat, $C(T, H) = \gamma(H)/T$, as a function of magnetic field for YBCO 6.56. Red circles are extracted from $C(T)$ measurements. The red oscillatory data are from the $C(H)/T$ trace at $T = 1K$ and the blue oscillatory data are from the $C(H)/T$ trace at $T = 2K$ after subtraction of well-characterized background terms (17) (18) (19) (20) . The cyan line is a fit to $A_c\sqrt{H}$ where the pre-factor is determined to be $A_c = 0.47\ mJmol^{-1}K^{-2}T^{-1/2}$. The black oscillation trace is $C(T = 1K, H) = A_C\sqrt{H} + AJ_0(\frac{4\pi t_\omega}{\hbar\omega_c})\cos\left(2\pi\left(\frac{F}{H} - \frac{1}{2}\right)\right)$ with the fitt parameters described in the text. The discrete data points (red circles) lie on top of the oscillations, showing that the increasing scatter with magnetic field comes from the oscillatory component of the specific heat. (B) The electronic component to the specific heat with the zero magnetic field contribution subtracted out and the oscillatory component subtracted out. (C) The data of Figure 1(A) with the p (H) subtracted out.

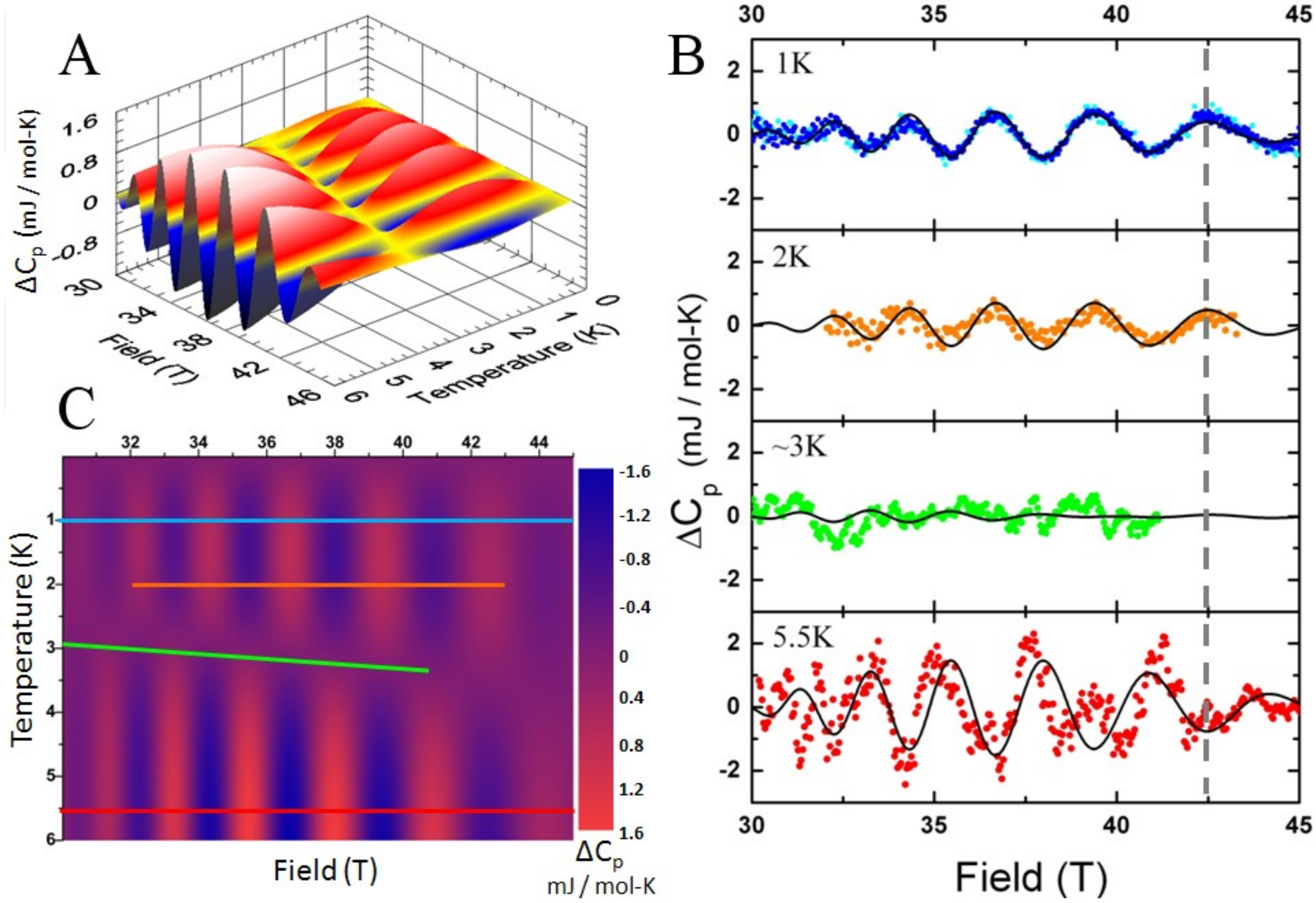

Figure 2: (A) Model of the quantum oscillatory phenomenon in specific heat using the LK formula and fitting parameters discussed in the text for a single $531 \pm 3T$ pocket with a warping of $15{:}2 \pm 0{:}5T$ . Note that the location of the node in the C(T,H) contour depends only on m*. For YBCO 6.56, we find m*= $1{:}34 \pm 0{:}06\ m_e$. (B) Oscillatory part of the specific heat data (scatter plots) with profile traces from the 3D contour (solid black) superimposed. For $1K$ the blue/cyan traces result from sweeping the magnetic field up/down between 30T and 45T and there is no observable hysteresis. The vertical dashed line shows _ phase shift between traces at $2K$ and $5{:}5K$. (C) Color density plot of (A). Colored lines indicate the locations in the temperature-field plane of the corresponding data (color) and profile trace (black) from (B).

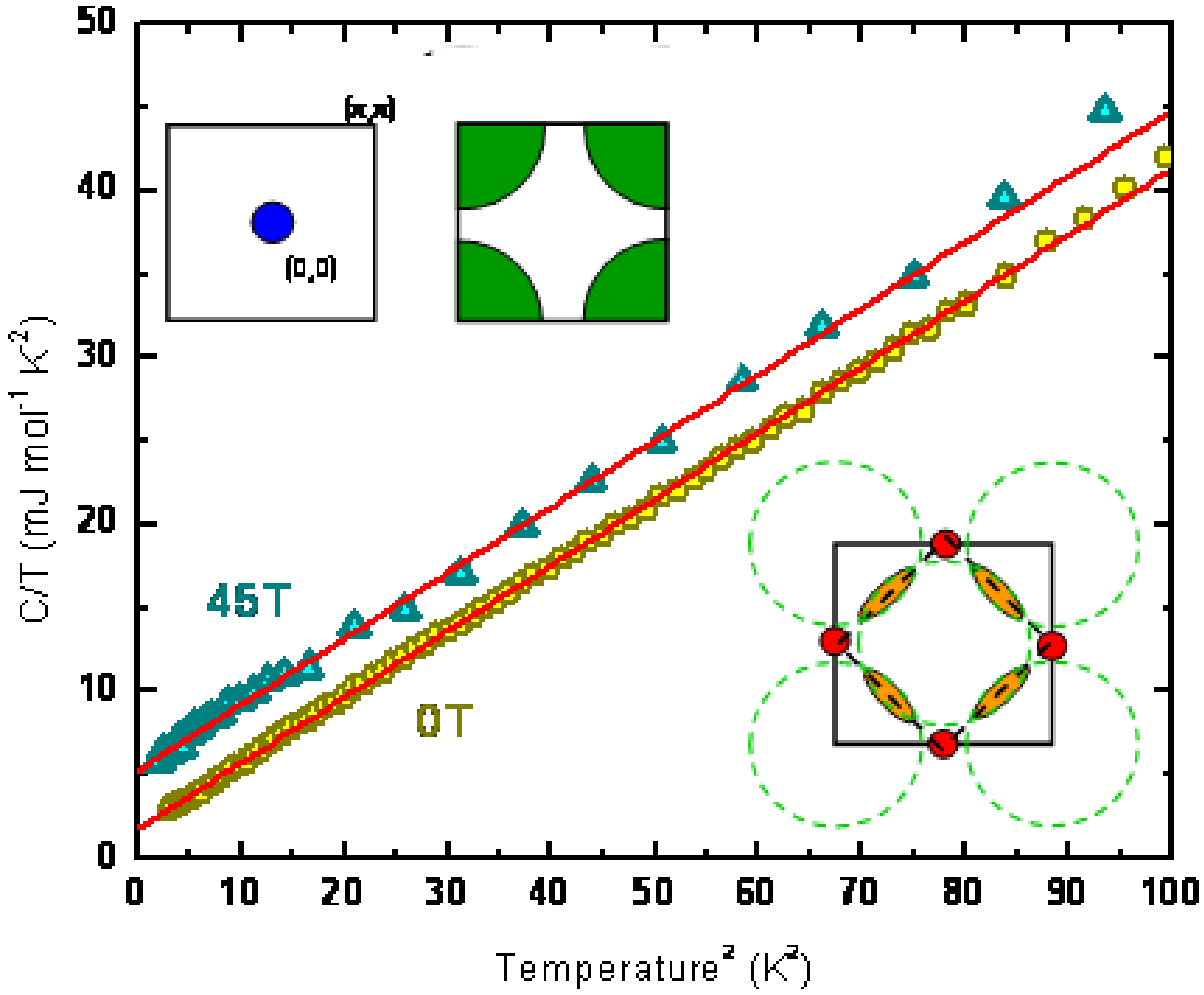

Figure 3: *C/T* as a function of $T^2$ for zero field (yellow circles) and the highest field measured, 45 T (blue triangles) with the field applied along the c-axis. In both the zero-field superconducting state and the high-field resistive state, the temperature dependence of the specific heat looks Fermi liquid like, as the only contributions appear to be from a (linear) electronic and (cubic) phonon term. The phonon term is field independent, and the slope $\beta = 0.395\ mJ\ mol^{-1}K^{-4}$, is consistent with previous reports (*18*) (*17*). The upper left insets are schematic Fermi surfaces for a single pocket centered at the (0,0) and ($\pi,\pi$) points. The lower right inset is a schematic representation of a Fermi surface reconstruction giving one electron pocket and two hole pockets per $CuO_2$ plane, resulting in $\gamma_{total} = 26\ mJ\ mol^{-1}\ K^{-2}$.

# Accompanying Material

## 1. Experimental determination of γ and $C(1K)/T$

Specific heat is a bulk thermodynamic measurement that probes all excitations in a system. In order to extract the excitations arising only from the electronic quasiparticle density of states, all other contributions to the specific heat need to be subtracted out. There are six contributions to the temperature and magnetic field dependence of the specific heat in YBCO:

$$C(T,B) = C_{electronic} + C_{phonon} + C_{Schottky} + C_{hyp} + C_{oscillatory} + C_{node}. \qquad (1)$$

The six terms represent, in order, delocalized electrons, lattice phonons, low -magnetic-field Schottky-like anamoly-currently attributed to unpaired spin- 1/2 paramagentic moments, hyperfine contribution arising from the interaction of nuclear magnetic moments with magnetic fields, the additional oscillatory component of the electronic specific heat upon sweeping magnetic field, and, finally, the temperature-dependent contribution of thermally excited quasiparticles in vicinity of d-wave nodes. For YBCO, $C_{node}$ is often hard to detect and early magnetic field studies of the specific heat have reported the best fit for $\alpha$ to be negative (*33*), which is non-physical. The difficulty arises in the very narrow range of temperature in which _ contributes any meaningful percentage of the total specific heat (*17*). The $C_{node}$ term does not need to be incorporated into our data analysis in order to obtain reliable and physically motivated fits. As such, the data analysis done here does not consider the expected zero field term from a gap with line nodes; $C_{node} = T^2$, where $\alpha$ characterizes the shape of the nodes. Since $\alpha$ scales with the slope of the line nodes in the d-wave superconducting gap, an estimate of $\alpha$ can be determined for YBCO 6.56 from thermal conductivity measurements (*29*). Previous measurements on YBCO 6.95 give $\alpha = 0.11$ $mJ\,mol^{-1}\,K^{-2}$ (*17*). Assuming an anisotropy from thermal conductivity measurements where, $\frac{\left(\frac{v_F}{v_\Delta}\right)_{6.99}}{\left(\frac{v_F}{v_\Delta}\right)_{6.54}} \sim 2$, would yield an $\alpha \sim 0.05$ $mJ\,mol^{-1}\,K^{-2}$ for YBCO 6.56 (*29*). This value of $\alpha$ is neglibible for extracting values for γ in YBCO 6.56.

The term describing the lattice phonon excitations, $C_{phonon} = \beta T^3$, is a field independent quantity where the pre-factor $\beta = \frac{2\pi^4 R}{5\Theta_D^3}$ and $\Theta_D$ is the Debye temperature. For the sample measured, the Debye temperature is 395 K. When global fits (see figure 9) are done to extract γ, and $\beta$ is allowed to be independent for each C(T) at fixed field values for $\beta$ range from $\beta = 0.392$ -0.420. The small changes in $\beta$ may be due to the $\alpha T^2$ term as described in ref. (*17*), or the interpolation procedures when corrected for the magnetoresistance of the cernox thermometer. Figure 4 shows the specific heat divided by temperature as a function of $T^2$ for zero and 45*T* magnetic fields. The slopes of these two lines are identical, where the high temperature data is fit to the following equation

$$C/T = \gamma + \beta T^2 \qquad (2).$$

The constant value for $\beta$ over the field ranges measured establishes β as a field independent quantity. In both the low-magnetic-field and very-high-magnetic field regimes, the only significant contribution to the specific heat at high temperatures are phonon excitations. The phonon contribution can now be reliably subtracted.

In the low and intermediate field ranges there is a "Schottky-like" contribution of the form $C_{Schottky} = \frac{Nz^2e^z}{(e^z+1)\ ^2}$ where $z = \frac{2gS\mu_B H_{eff}}{k_B T}$$C_{Schottky}$ with $H_{eff} = \sqrt{H_{applied}^2 + H_{internal}^2}$ (*17*) and it needs to be subtracted out. In the high temperature regime where $k_bT << \Delta_e$ the specific heat has a temperature dependence where, $C \propto \frac{e^{\frac{1}{T}}}{T^2}$ giving way to exponential activation. In the high temperature regime $kbT >> \Delta_e$ and specific heat energy scale, $C \propto \frac{H_{eff}^2}{T^2}$. This accounts for the "upturn" at low temperatures. These limiting behaviors work for any $n > 1$ level system. An applied magnetic field has the effect of decreasing the amplitude, moving the peak up in temperature and increasing its width in temperature. At very high fields the Schottky term becomes negligible as it merges into the background. The nature of the Schottky term is not well understood (*32*) and is currently ascribed to residual paramagnetic centers associated with unpaired copper spin-1/2 moments (*17*). The standard treatment of this zero-field upturn at low temperatures is to assume a small internal field caused by disorder, as the strength of the effect increases with increasing disorder (*32*). In zero field, $H_{eff}$ must be small since the moments are ordered only by internal interactions. While the data can be modeled well with this formalism, $H_{eff}$ is a fit parameter which is itself a function of the applied field and temperature. Thus, we analyze the data two ways in regards to the low-magnetic-field electronic Schottky anomaly. First, the traditional manner in which $H_{internal}$ is allowed to be a fit parameter over the range of 0T-16T where the Schottky-like anomaly plays a role; and, second, all data taken for YBCO 6.56 assume $H_{internal} = 0$ making $H_{eff} = H_{applied}$ removing it as a fit parameter. Fits to the specific heat are done globally over four different field regions which have physical meaning. We now treat these regions systematically.

The low-field regime is defined as the field range in which only the high temperature limit of the electronic Schottky term contributes and the total specific heat takes the form

$$C_{LFR} = C_{electronic} + C_{HighT\ Schottky} = \gamma T + Nz^2. \qquad (3)$$

γ and $C(1K)/T$ can now be determined in low fields by two independent fitting procedures. First a linear fit over only the high temperature region of the data in Figure 4 is determined and then fixed, leaving only one remaining fit parameter, γ. The second treatment is shown in Figure 5 in which equation 5 is used to globally fit the family of curves over the entire temperature range. The resulting values for γ are consistent and plotted in Figure 9.

The intermediate-field regime is defined when the full form for the Schottky term must be included in the specific heat analysis.

$$CIFR = C_{electronic} + C_{Schottky} = \gamma T + \frac{Nz^2e^z}{(e^z+1)^2} \qquad (4)$$

Figure 6 shows the field range of $2T$ - $8T$. The $2T$ data are fit using both low and intermediate-field protocols as a consistency check. From equation 5, the only parameters which are allowed to float in the fit are γ and $N$. $z$ is a global fit parameter which gives the best fit of $z = \frac{2gS\mu_B H_{applied}}{k_B T} \approx 1.37 \frac{H_{applied}}{T}$ which is close to the experimentally determined value from muon experiments, $gS \approx 1{:}03$ (?). When the field value reaches $8T$, equation 6 no longer works well, as the nuclear Schottky anomaly begins to play a role at lowest temperatures. In the high-field regime $H \approx 6T$-$15T$ , the Schottky-like anomaly can easily be observed by following its quick

suppression as field increases. The high-field regime and intermediate field protocols are both used in the 6$T$ and 8$T$ traces where there is a small but non-negligible nuclear Schottky term.

$$C_{HFR} = C_{electronic} + C_{Schottky} + C_{hyp} = \gamma T + \frac{Nz^2 e^z}{(e^z+1)^2} + \frac{P\Delta_{hyp}^2 H_{applied}^2}{T^2} \qquad (5)$$

Where P is the number of nuclear spins. Equation 7 has the largest number of fit parameters. The low-field Schottky-like behavior is suppressed at such a fast rate that small changes in the fit parameters can have a large effect on γ, as the $C_{Schottky}$ term still has an exponential and power law response. Global fitting of the data at fields above and below this regime are important as they provide constraints for the fit parameters. High field measurements therefore constrain $\Delta_{hyp}$, while low and intermediate fields constrain *z*, leaving only *N* and γ as fit parameters. Once high enough fields are reached, $H \approx 20T$, $\Delta_{hyp}$ is found to essentially be constant. At lower fields, the full value of $\Delta_{hyp}$ is not yet reached.

Further complicating the high field regime is the effect of magneto-resistance in the calorimeter. At 20$T$ the magneto-resistance of cernox thermometers reach a local maximum and changes sign (*?*) (*?*) implying that any calibration noise will be amplified in the 15$T$ - 25$T$ regime.

The very-high-field regime is defined by the vanishing of the low temperature Schottky anomaly such that it no longer contributes to the specific heat ($H > 20T$). In this regime the only subtraction needed to determine the electronic contribution comes from the $C_{hyp}$, the interaction of copper nuclear magnetic moments with the applied magnetic field (*32*). As the copper

moment is several orders of magnitude smaller than the electronic moment, the hyperfine density of states contribution will only play a role at high fields and low temperatures.

$$C_{VHFR} = C_{electronic} + C_{hyp} = \gamma T + \frac{P\Delta^2_{hyp}H^2_{applied}}{T^2} \qquad (6)$$

The very-high-field regime fit is much like that of the low-field regime since one can fit the high temperature data using equation 4. Equation 8 can also be used where the second term is caused by the hyperfine interaction, and, the only fit parameter is γ, as $\Delta_{hyp}$ can be globally determined and found to be nearly constant with magnetic field using both C(T) for different fields and C(B) at fixed temperature.

The different protocols for fitting the C(T) as a function of magnetic field data are shown in Figure 9. Verifying the $\sqrt{H}$ dependence of γ using a plethora of protocols demonstrates the robust nature of the effect.

The 0*T*-16*T* traces were done in an Oxford cryogenic magnet in a home built He-3 system. The 30*T,* 40*T,* 45*T,* and *C*(*H*) measurements were made in the 45T Hybrid magnet at the National High Magnetic Field Laboratory in Tallahassee Florida. A He-4 system was used for measurements from 1.5*K* to 10*K* for 20*T,* 20*:*5*T,* 22*T,* 24*T,* and 28*T* fields.

## 2. Extracting Background information from the *C*(*H*) quantum oscillations

For very high magnetic fields, it is possible to generalize the nuclear Schottky anomaly due to the copper nucleus with a spin-3/2 system. The high temperature limit exhibits the same $\Delta\frac{\Delta H^2}{T^2}$, dependence for all values of spin. After subtraction of the phonon term, only the high temperature part of the nuclear Schottky term, needs to be subtracted. The data are therefore fit in the low temperature regime $C(T,H = 40T\,)/T$ to the following equation

$$\frac{C(T,40T)}{T} = \gamma + \frac{H_T}{T^3} \qquad (7)$$

where $H_T = \Delta_2 H^2 = 18.7 mJmol^{-1}K^{-2}$ is the Schottky coefficient at fixed magnetic field. Next we hold the temperature constant and examine the magnetic field dependence of the specific heat, best fit to a modified form from equation [14] in ref (*20*)

$$\frac{C(1K,H)}{T} = \gamma(H, T = 1K) + \ H_H H^2 + A_c\sqrt{H} + \alpha T \qquad (8)$$

where $\gamma(0)$ is $2mJ{=}(mol/K^2)$ in our case, $\alpha$ is negligible and set to and $H_H$ is the nuclear Schottky coefficient at fixed temperature. Comparing the two data sets provides a consistency check.

$$H_H T^3 = 0.0108(1)^3 = \Delta^2 \rightarrow \Delta = 0.104 \qquad (9)$$

$$\frac{H_T}{H^2} = \frac{18.7}{40^2} = 0.0117 = \Delta^2 \rightarrow \Delta = 0.108 \qquad (10)$$

We find that the values for the nuclear Schottky energy gap are the same from both C(T) in fixed magnetic field, and C(H) at fixed temperature. Also, the $\sqrt{H}$ pre-factor, $A_c = 0.434$, which closely matches the value of $A_c = 0.464$ determined from Figure 9. With these consistency checks we find that the quantum oscillations sit on the $\sqrt{H}$ background. Interestingly, the specific heat at high magnetic fields can be fit to a formula which includes only the electronic, nuclear Schottky, and d-wave gap contributions.

## 3. Electronic Schottky anomaly

The electronic "Schottky-like" anomaly in cuprates is attributed to chemical impurities on the copper sites that induce a local moment (*32*). A simple example is the case of Zn or Cr substitution onto Cu sites where Zn actually removes a Cu spin-1/2 moment. Specific heat on Cr substituted $YBa(Cu_{1-y}Cr_y)_3O_7$ has a low field contribution that is well modeled by a two level Schottky system with a g-factor of 2 (*32*). These Zn and Cr specific heat measurements are the motivating factor, for attributing the electronic Schottky term in YBCO as paramagnetic centers arising from disorder (*32*).

The authors would like to thank Neil Harrison, Ross MacDonald,

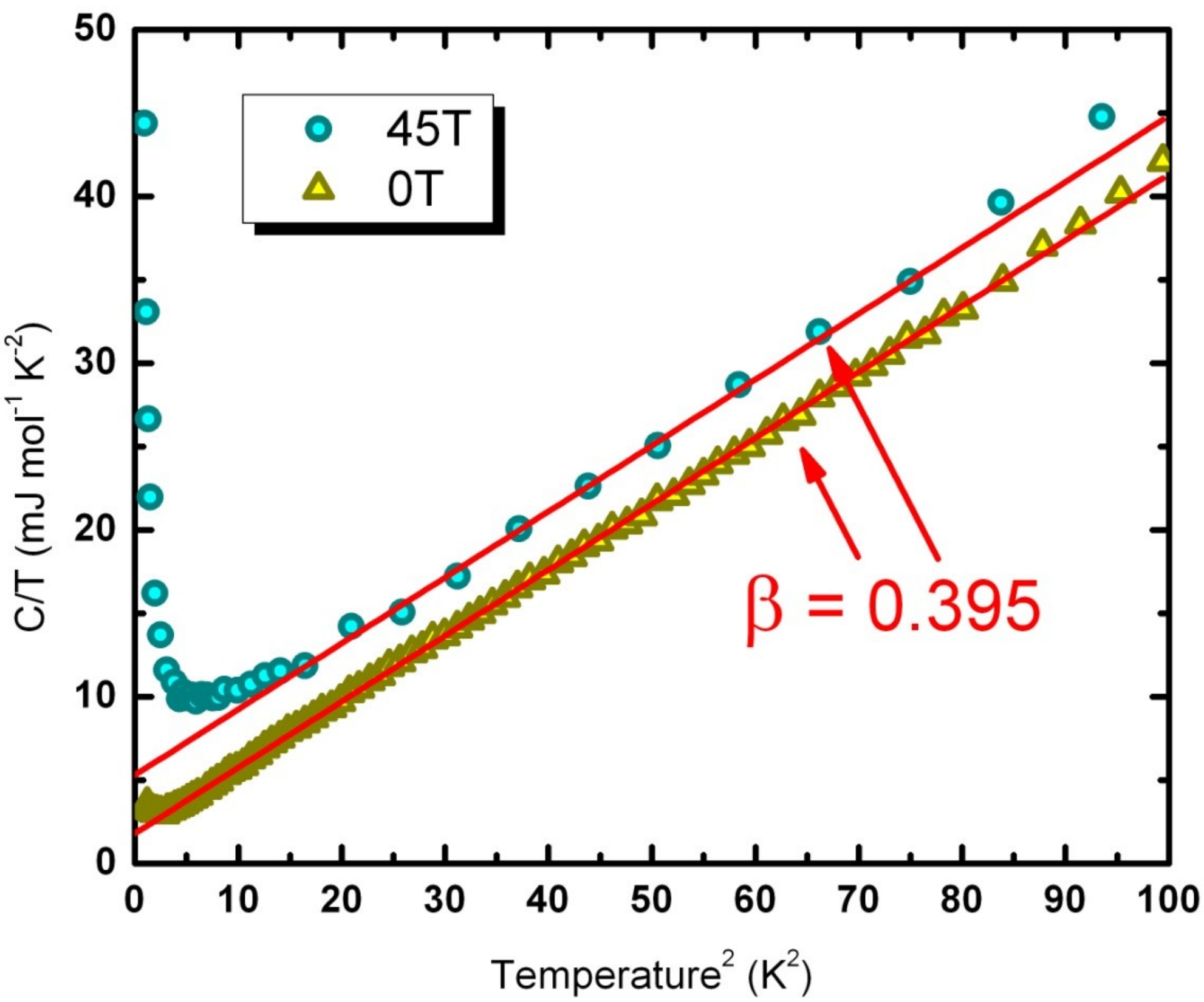

Figure 4: *C/T* as a function of $T^2$ for zero field (yellow triangles) and the highest field measured, 45T (cyan circles) with the field applied parallel the c-axis. For this sample β = 0*:*395 *mJmol$^{-1}$K$^{-4}$* while (*19*) found β = 0*:*392 *mJ mol$^{-1}$K$^{-4}$* and (*17*) found β = 0*:*392 *mJ mol$^{-1}$K$^{-4}$*. In both of these regimes the specific heat looks Fermi liquid like as, other than the low-temperature upturn due to the Schottky anomaly, the only contributions appear to be from a phonon and electronic term. The phonon term is field independent as shown by the same value of the linear slope for all fields measured.

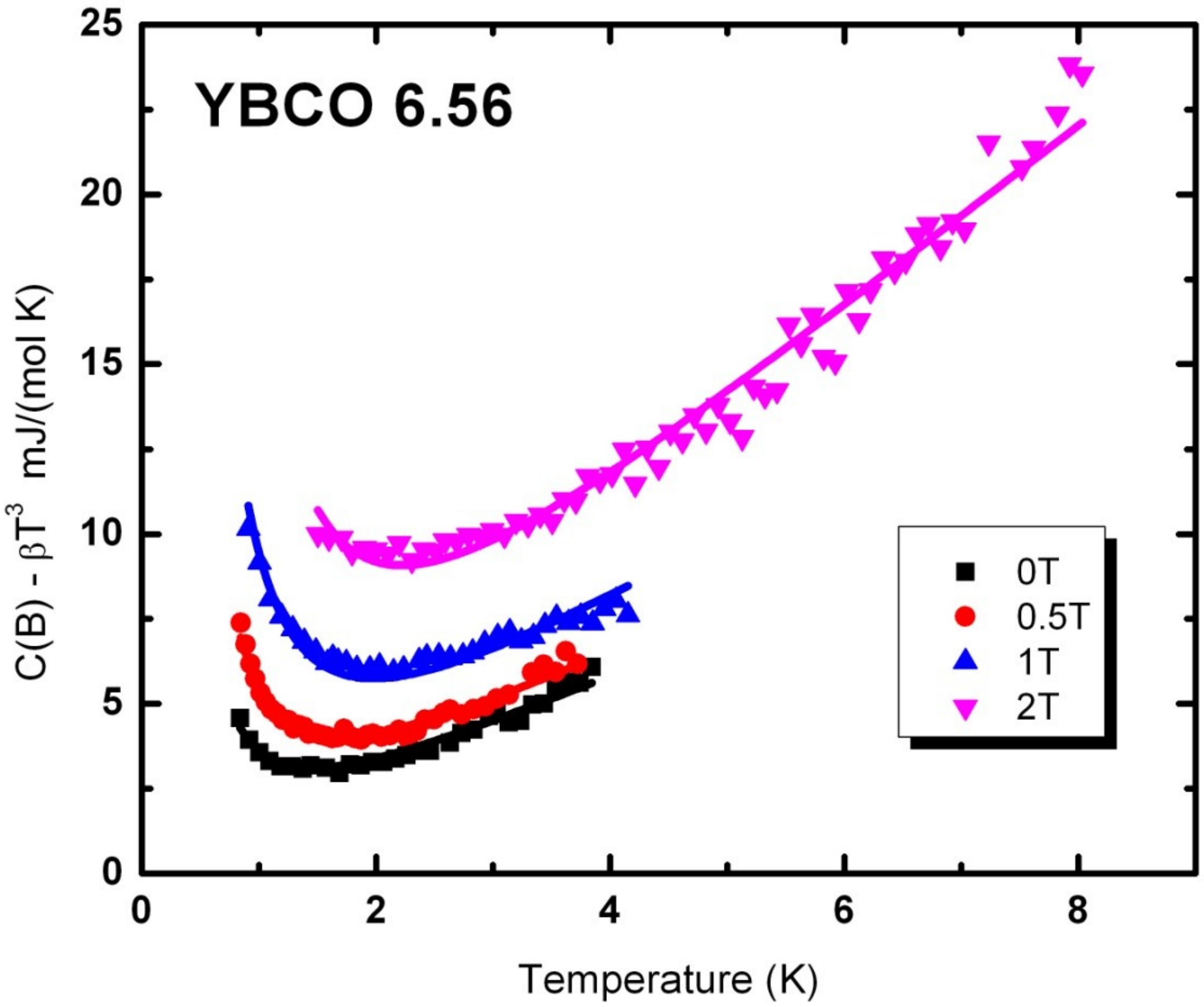

Figure 5: Specific heat with the phonon contribution subtracted out, as a function of temperature in the low-field regime. The low-field determination of γ is fit using equation 5 where with $H_{eff} = H_{applied}$ which captures the Schottky-like behavior at low temperatures. The zero-field data is shown but not included in the global fit when $H_{eff} = H_{applied} = 0T$.

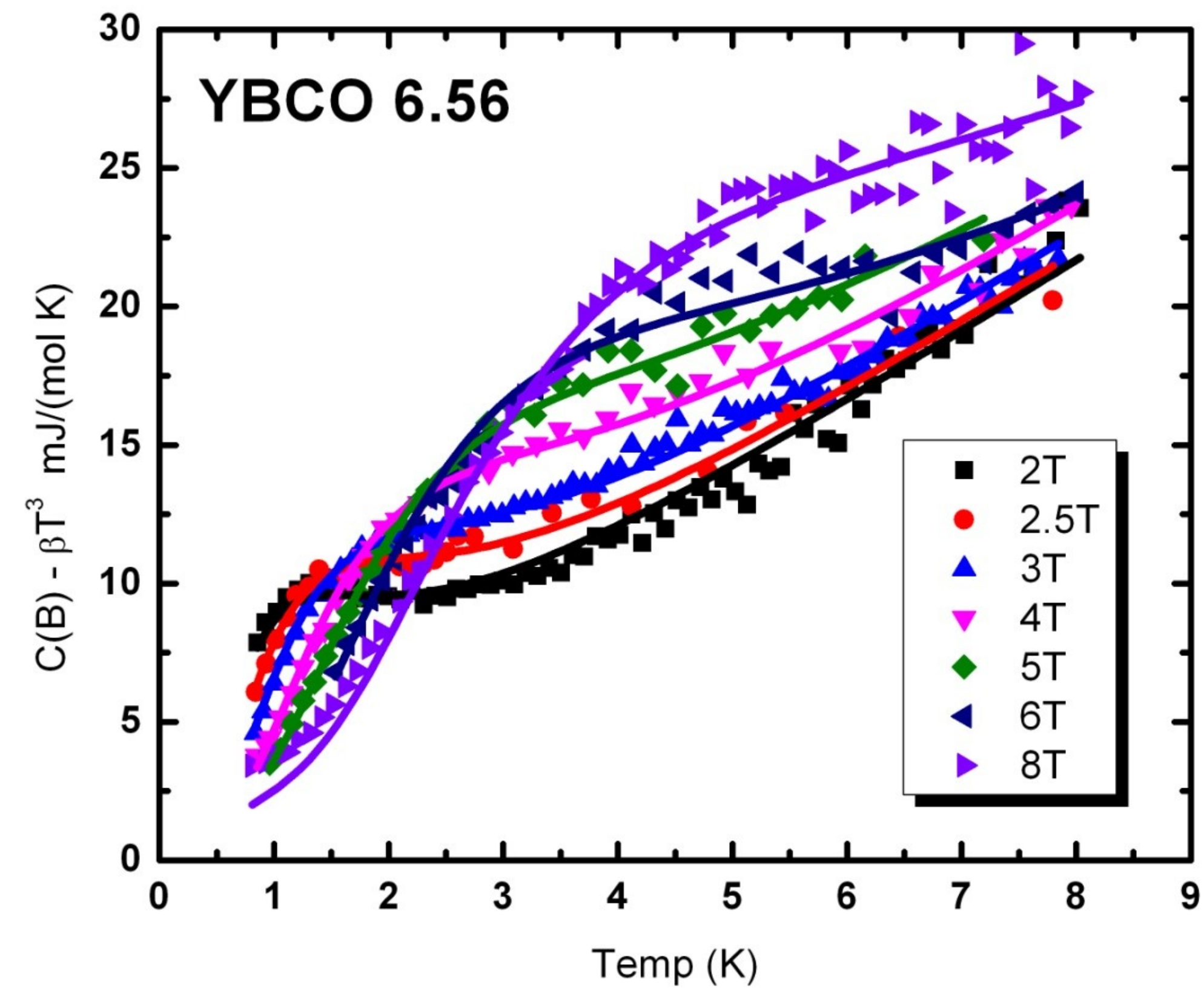

Figure 6: Specific heat as a function of temperature in the intermediate-field range. The Schottky-like contribution is largest in this field regime. At 8T the Hyperfine term at the lowest temperature accounts for the upward deviation of the data from the equation 6 fit. The addition of the hyperfine term is the signature of moving from the intermediate-field regime to the high field regime (equation 5).

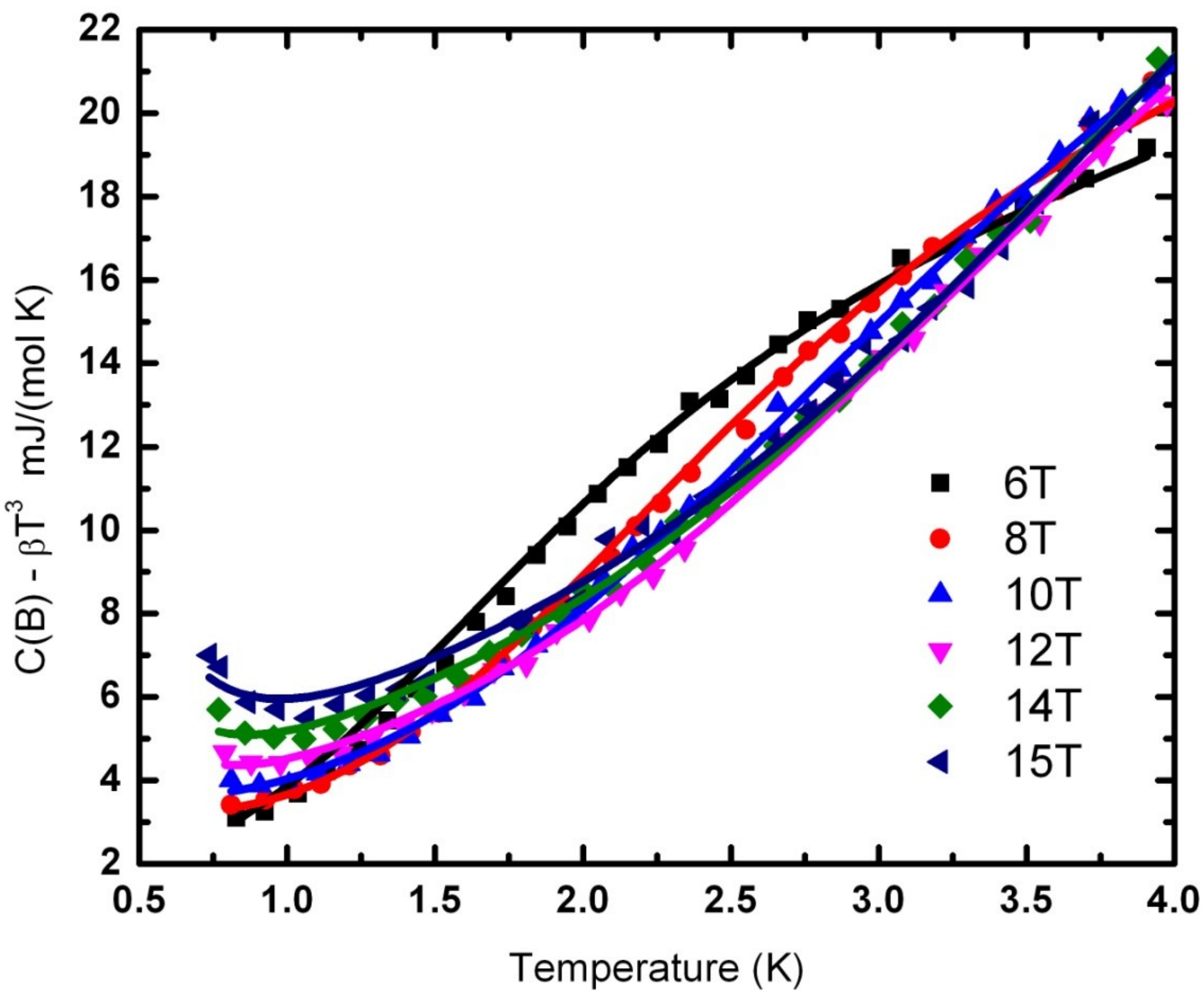

Figure 7: High-field specific heat with the phonon contribution subtracted out, as a function of temperature. Fits in the high-field regime, $\sim 6T - 16T$ , are the most difficult to fit, as the Schottky-like anomaly at $T \sim 2K$ is quickly suppressed from $6T$ to $14T$ while the hyperfine contribution at low temperature grows rapidly with increasing field

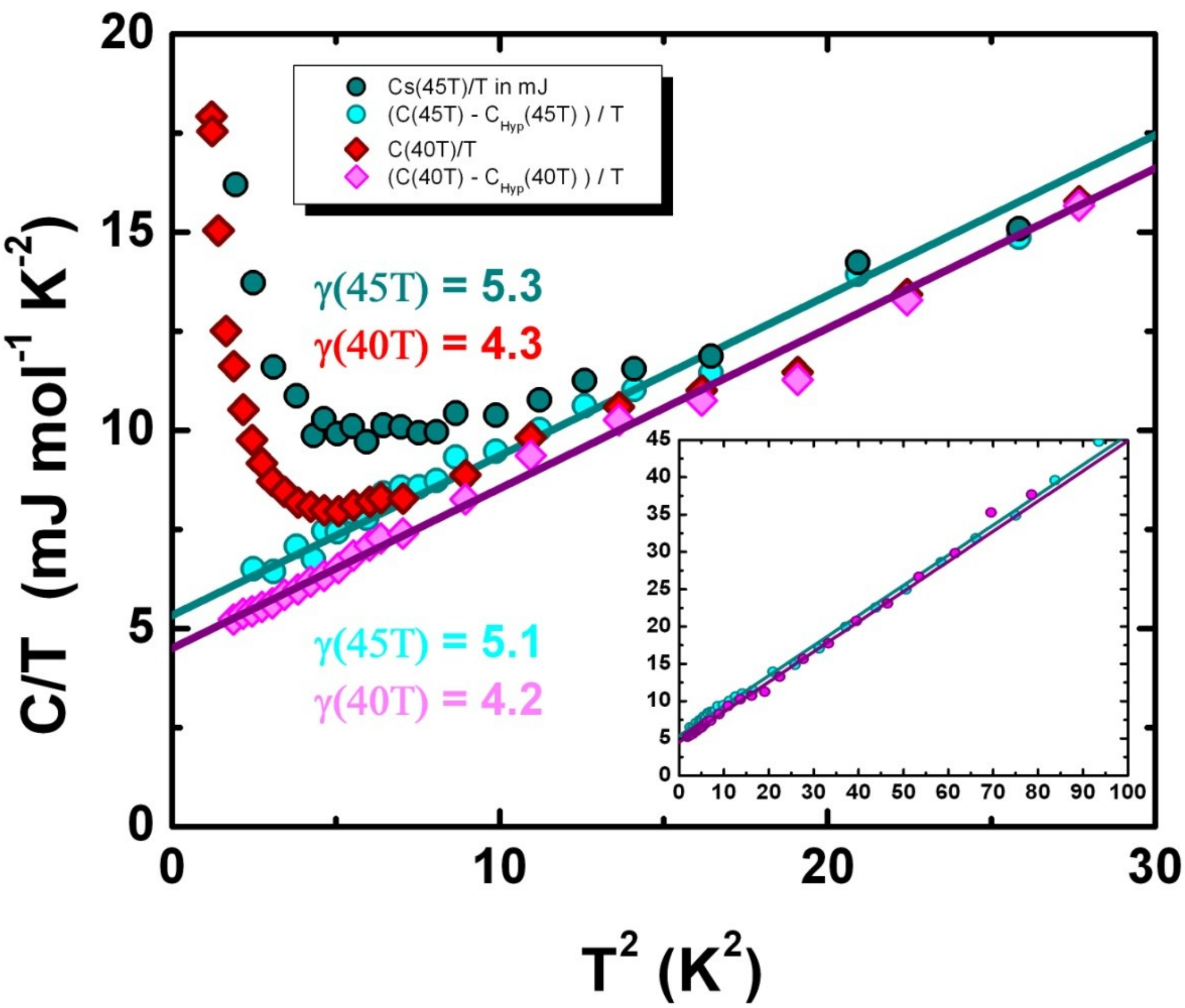

Figure 8: *C/T* as a function of $T^2$ for 40*T* and 45*T*. In the very-high-field limit the only contribution to the specific heat that must be subtracted is the hyperfine contribution. γ is the only fit parameter and can be calculated by either equation 4 or equation 8. The results to the fits are shown in the inset, and are independent the fitting protocol.

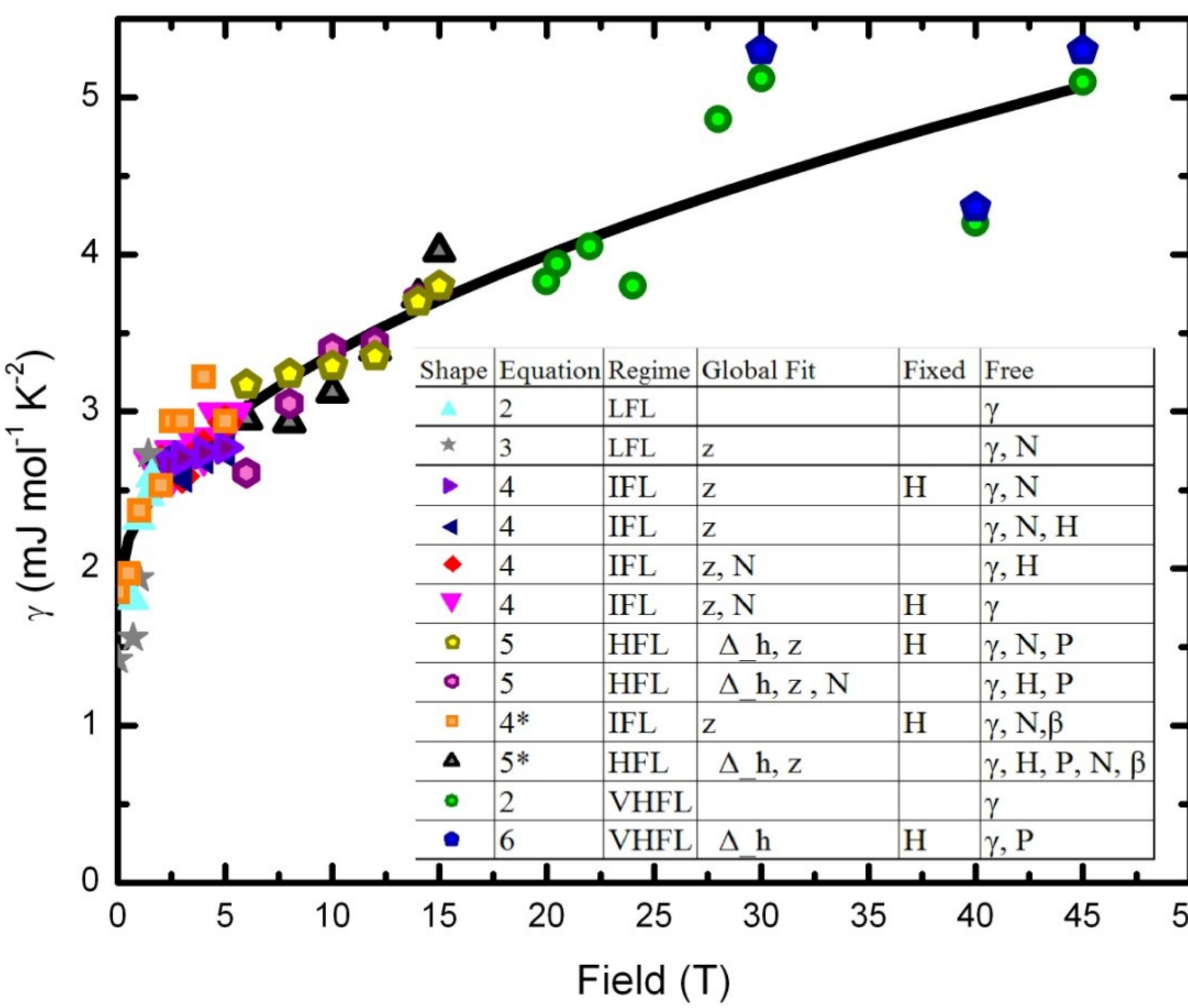

| Shape | Equation | Regime | Global Fit | Fixed | Free |
|---|---|---|---|---|---|
| ▲ | 2 | LFL | | | γ |
| ★ | 3 | LFL | z | | γ, N |
| ▶ | 4 | IFL | z | H | γ, N |
| ◀ | 4 | IFL | z | | γ, N, H |
| ◆ | 4 | IFL | z, N | | γ, H |
| ▼ | 4 | IFL | z, N | H | γ |
| ⬟ | 5 | HFL | Δ_h, z | H | γ, N, P |
| ⬢ | 5 | HFL | Δ_h, z , N | | γ, H, P |
| ■ | 4* | IFL | z | H | γ, N,β |
| △ | 5* | HFL | Δ_h, z | | γ, H, P, N, β |
| ● | 2 | VHFL | | | γ |
| ⬟ | 6 | VHFL | Δ_h | H | γ, P |

Figure 9: γ calculated by equations stated in the text and figures, plotted as a function of applied magnetic field. The black curve is a $\gamma(H) = A_c\sqrt{H}$ fit through all the data with $A_c = 0.47$ $mJmol^{-1}K^{-2}$. The inset details the fitting constraints on the fit parameters. The equations labeled with a star (*) refer to the equation with the phonon $\beta T^3$ term added. For protocols with fixed $H$, $H_{eff} = H_{applied}$. The table inset shows 12 protocols for treating the data.

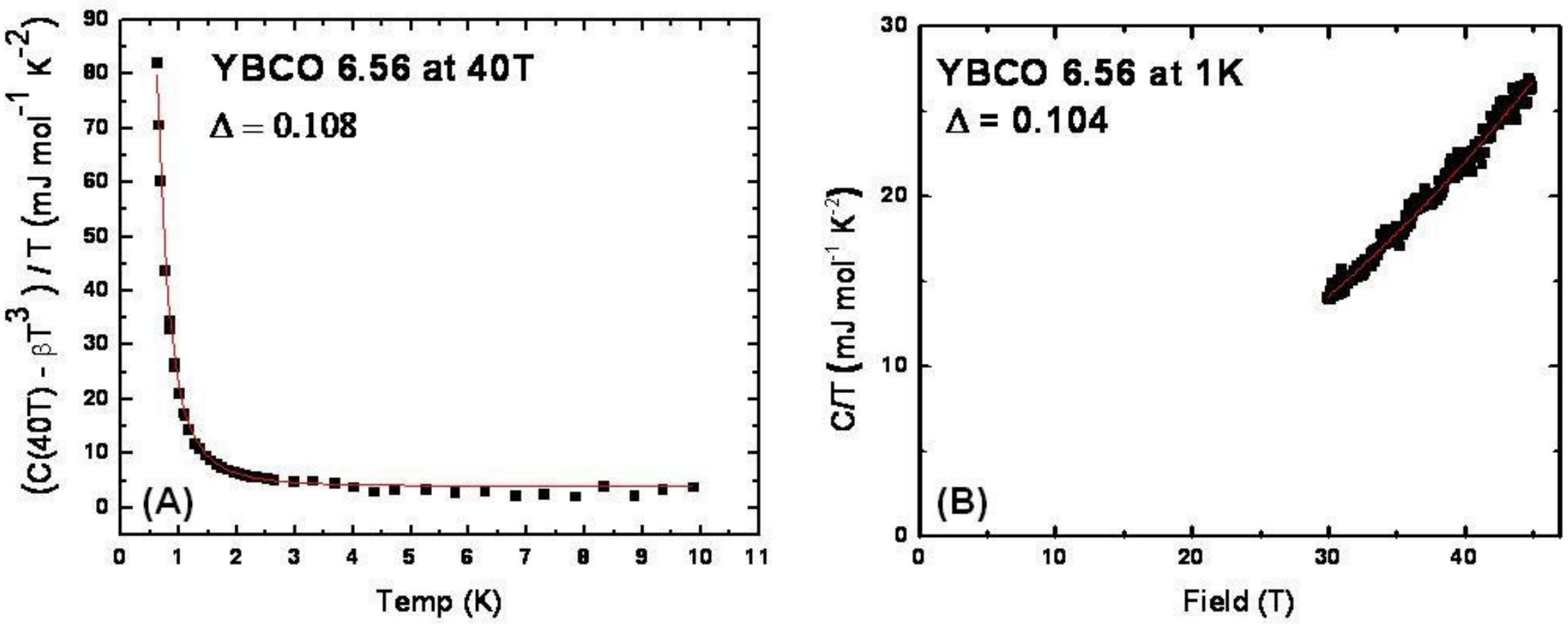

Figure 10: (A) Specific heat at $40T$ divided by temperature as a function of temperature for YBCO 6.56 with the phonon contribution removed (black squares). The red curve is a fit in to equation 9 where the only contributions are the electronic and high temperature limit of the nuclear Schottky. The fit parameters for $H_T = \Delta^2 \frac{H^2}{T^3} = \Delta^2 (40)^2 = 18.7$. (B) $C(1K)/T$ as a function of magnetic field for YBCO 6.56 (black circles). The red line is a fit to data using equation 10 with γ(0) fixed from zero field $C(T)$ measurements. The values for $A_c = 0.434$ and the nuclear Schottky term $\Delta^2 = 0.0108T^3$ closely match the values determined from $C(T)$ fixed field measurements where $A_c = 0.46$. Note both data taken swept field at fixed temperature, and swept temperature at fixed field give a common value for $\Delta = 0.106 \pm 0.002$

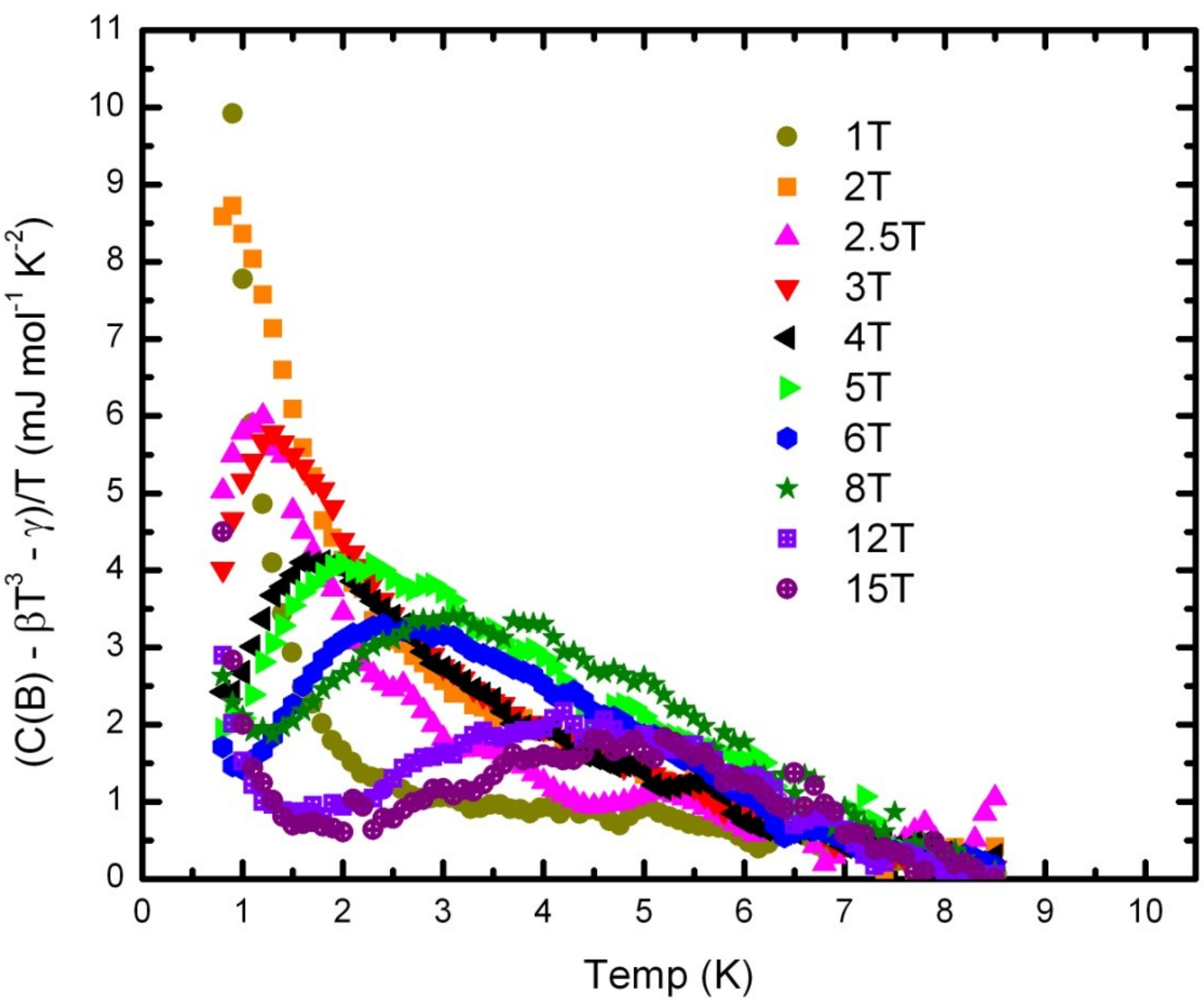

Figure 11: Raw data showing the electronic Schottky contribution as a function of temperature at varying magnetic fields.